\documentstyle[11pt,epsfig]{article}
\def\edth{\;\raise1.0pt\hbox{$'$}\hskip-6pt\partial\;}
\def\baredth{\;\overline{\raise1.0pt\hbox{$'$}\hskip-6pt
\partial}\;}
\def\gsim{~\rlap{$>$}{\lower 1.0ex\hbox{$\sim$}}}

\def\be{\begin{equation}}
\def\ee{\end{equation}}
\def\ba{\begin{eqnarray}}
\def\ea{\end{eqnarray}}
\def\h{\rm{h}}

\def\B{\rm{B}}
\def\DM{\rm{DM}}

\def\h{\rm{h}}
\def\i{\rm{i}}
\def\de{\rm{de}}
\def\dec{\rm{dec}}

\newcommand{\fr}[2]{\frac{#1}{#2}}
\newcommand{\ws}{\rm{ws}}
\newcommand{\sw}{\rm{sw}}
\newcommand{\omde}{\omega_{\rm{de}}}
\newcommand{\Omo}{\Omega_{\rm{m}}^{0}}
\newcommand{\Odeo}{\Omega_{\rm{de}}^{0}}

\begin{document}

\title{Dark matter growth and baryon bias in an accelerating universe}

\author{Seokcheon Lee$^{\,1,2}$}

\maketitle

$^1${\it Institute of Physics, Academia Sinica,
Taipei, Taiwan 11529, R.O.C.}

$^2${\it Leung Center for Cosmology and Particle Astrophysics, National Taiwan University, Taipei, Taiwan 10617, R.O.C.}

\begin{abstract}
We investigate the exact analytic solutions for the growths of the dark matter and the baryon in sub-horizon scale. The growth of the dark matter $\delta_{\DM}$ is related to that of the halos. Thus, the exact solution for the growth of the dark matter is important to obtain the proper properties of dark matter halos. However, the dark energy model dependence of $\delta_{\DM}$ is confused with the $\delta_{\DM}$ dependence on $\Omega_{m}^{0}$. Thus, the careful investigation is necessary for the $\delta_{\DM}$ dependence on dark energy models. We also obtain the exact solution of the growth of the baryon $\delta_{\B}$ which can be used to obtain the baryon bias factor $b(a)$. This might be able to be observed in intracluster gas or in Lyman-$\alpha$ clouds. However, $b(a)$ is quite model independent. Recently, we obtained the exact analytic solution for the growing mode solution of the matter linear density perturbation $\delta$ on sub-horizon scale for general dark energy model \cite{SK}. This solution is not same as the well known approximate analytic solution \cite{Waga}. The exact analytic solution shows the same evolution behavior of the growth factor obtained numerically. However, the exact solution is simple and useful for the extension to other models including modified gravity theories. Furthermore, it guides to the fact that the growth index parameter depends on both models $\omega_{\de}$ and $\Omega_{m}^{0}$ and thus we need to be careful for applying the fitting formulae to the general models \cite{WS}. The exact analytic solutions for the growth factor will provide the more accurate tools for the weak lensing, the number density of clusters, their mass and etc.

\end{abstract}

In the concordance model of cosmology baryon contributes about $4 \%$ of the total energy density with  $23 \%$ cold dark matter and $73 \%$ dark energy in the universe \cite{concordance}. The linear perturbation of the dark matter is related to the halos density perturbation. Thus, the exact solution for the dark matter growth $\delta_{\DM}$ provides the accurate and powerful tool for the probing dark matter halo properties. Also after the decoupling from the photon, the linear perturbation of the baryon $\delta_{\B}$ grows in the gravitational potentials created by the dark matter. Zero pressure of dark matter causes the uniform growth of the dark matter and the baryon in sub-horizon scale. Thus, the growth of baryon is related to the growth of the dark matter through the baryon bias factor $\delta_{\B} = b(a) \delta_{\DM}$ \cite{Padmanabhan}. The baryon bias is able to be observed when we consider not only the baryon distribution clustered in luminous objects but also the baryon not collapsed in galaxies.

We use the flat Friedmann-Robertson-Walker universe to probe the linear density perturbations   $\delta_{\DM}$ and $\delta_{\B}$ \ba H^2 \equiv \Bigl(\fr{\dot{a}}{a}\Bigr)^2 &=& \fr{8\pi G}{3}(\rho_{\DM} + \rho_{\B} + \rho_{de}) = \fr{8 \pi G}{3} \rho_{cr} \, , \label{H} \\ 2 \fr{\ddot{a}}{a} + \Bigl(\fr{\dot{a}}{a}\Bigr)^2 &=& - 8 \pi G \omega_{de} \rho_{de} \, , \label{dotH} \ea where $\omega_{de}$ is the equation of state (eos) of dark energy, $\rho_{cr}$ is the critical energy density, $\rho_{\DM}$, $\rho_{\B}$ and $\rho_{de}$ are the energy densities of the dark matter, the baryon and the dark energy, respectively. We consider the constant $\omega_{de}$. The sub-horizon scales linear perturbation equations for the coupled system of the dark matter and the baryon with respect to the scale factor $a$ is given by \cite{Padmanabhan, Bonnor} \ba \fr{d^2 \delta_{\DM}}{da^2} + \Biggl( \fr{d \ln H}{d a} + \fr{3}{a} \Biggr) \fr{d \delta_{\DM}}{d a} &=& \fr{4 \pi G \rho_{\DM}}{(aH)^2} \delta_{\DM} \, , \label{dadeltaDM} \\ \fr{d^2 \delta_{\B}}{da^2} + \Biggl( \fr{d \ln H}{d a} + \fr{3}{a} \Biggr) \fr{d \delta_{\B}}{d a} &=& \fr{4 \pi G \rho_{\DM}}{(aH)^2} \delta_{\DM} \, , \label{dadeltaB} \ea where we use the fact that the dominant contribution to the gravitation potential comes from the dark matter ({\it i.e.} $ \Omega_{\DM} \delta_{\DM} + \Omega_{\B} \delta_{\B} \simeq \Omega_{\DM} \delta_{\DM}$). We only consider the matter perturbation in Poisson equation because the dark energy is supposed to be smooth on sub-horizon scale \cite{Dave}. Recently, we obtain the exact analytic solution of sub-horizon scale growth factor of the matter with the general dark energy models \cite{SK}. Compared to the previous results in Ref. \cite{SK}, we just consider the dark matter as the source of the gravitational potential. Thus, the solution for the sub-horizon scale linear perturbation of the dark matter $\delta_{\DM}$ is almost identical with that of total matter given in the reference \cite{SK}
\ba \delta_{\DM}(Y) &=& c_{1} Y^{\fr{3 \omega_{de} -1}{6 \omega_{de}}} F [\fr{1}{2} - \fr{1+ \sqrt{1 + 24 R}}{12\omega_{de}}, \fr{1}{2} - \fr{1 - \sqrt{1 + 24 R} }{12 \omega_{de}}, \fr{3}{2} - \fr{1}{6 \omega_{de}}, -Y] \nonumber \\ && \, + \, c_{2} F[\fr{1- \sqrt{1 + 24 R} }{12\omega_{de}}, \fr{1 + \sqrt{1 + 24 R} }{12 \omega_{de}}, \fr{1}{2} + \fr{1}{6 \omega_{de}}, -Y] \,  \label{deltaDME} \\ &\simeq& c_{1} Y^{\fr{3 \omega_{de} -1}{6 \omega_{de}}} F [\fr{1}{2} - \fr{1}{2\omega_{de}}, \fr{1}{2} + \fr{1}{3 \omega_{de}}, \fr{3}{2} - \fr{1}{6 \omega_{de}}, -Y] \nonumber \\ && \, + \, c_{2} F[-\fr{1}{3\omega_{de}}, \fr{1}{2 \omega_{de}}, \fr{1}{2} + \fr{1}{6 \omega_{de}}, -Y] \, , \label{deltaDM} \ea where $Y = \fr{\Omega_{m}^{0}}{\Omega_{\de}^{0}} a^{3\omega_{\de}} \equiv Q a^{3 \omega_{\de}}$, $\Omega_{m}^{0} = \Omega_{\DM}^{0} + \Omega_{\B}^{0}$ and $R = \fr{\Omega_{\DM}^{0}}{\Omega_{m}^{0}}$. We use $R \rightarrow 1$ ({\it i.e.} $\Omega_{\DM}^{0} \rightarrow \Omega_{m}^{0}$) in the second approximation. Thus, the second equality is valid as long as the dark matter energy density dominates the baryon energy density. We use $\Omega_{\DM}^{0} = 0.23$, $\Omega_{\B}^{0} = 0.4$ and $\Omega_{\de}^{0} = 0.73$ later to be consistent with the concordance model. The baryon contribution is still about $17 \%$ of the dark matter. Thus, we use the solution of $\delta_{\DM}$ given in Eq (\ref{deltaDME}). As in our previous work \cite{SK}, we need to determine the coefficients of $\delta_{DM}$ to make the above analytic solution as a growing mode one. We obtain them from initial conditions of growing mode solution \be \delta_{\DM}(a_i) \simeq a_{i} \hspace{0.2in} {\rm and} \hspace{0.2in} \fr{d \delta_{\DM}}{da} \Bigl|_{a_{i}} = 1 \, . \label{ini} \ee
\begin{center}
\begin{figure}
\vspace{1.5cm}
\centerline{
\psfig{file=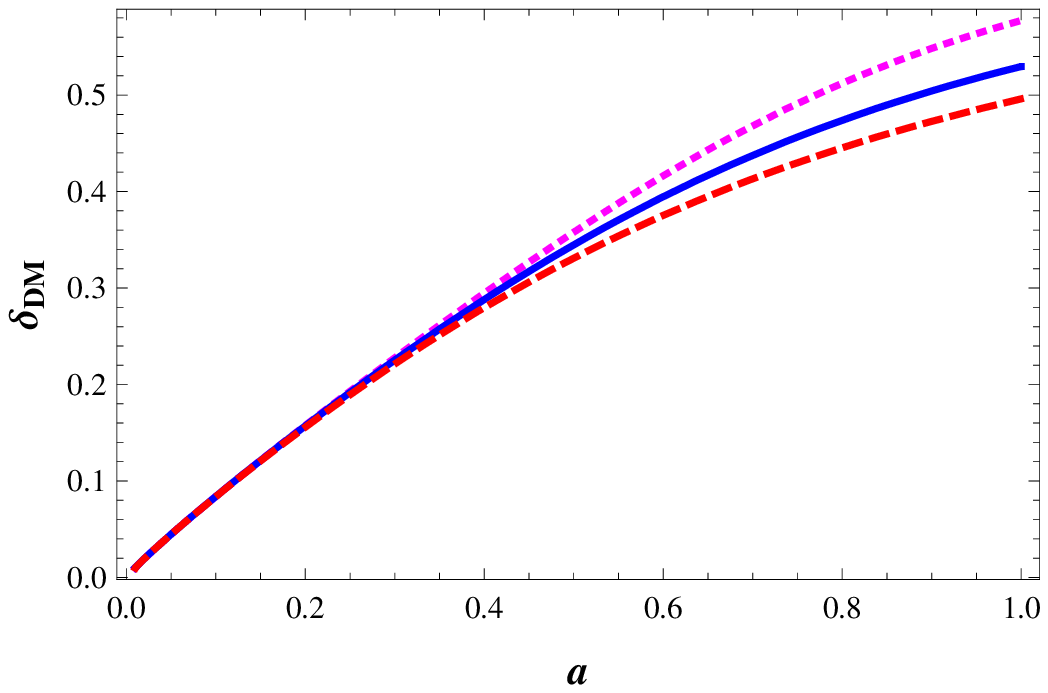, width=6cm} \psfig{file=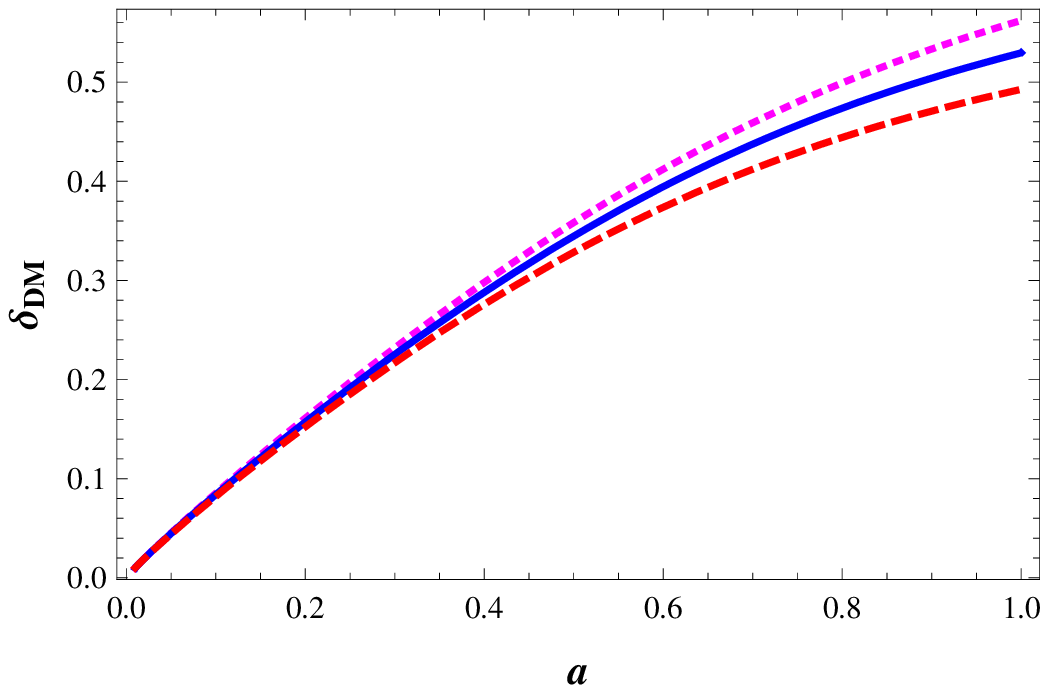, width=6cm} }
\vspace{-0.2cm}
\caption{ a) Evolutions of $\delta_{\DM}(a)$ when $\Omega_{\DM}^{0} = 0.23$ for $\omega_{de} = -1.5, -1.0$, and $-0.8$, respectively (from top to bottom).  b) $\delta_{\DM}(a)$ evolutions for the different values of $\Omega_{\DM}^{0} = 0.26, 0.23$, and $0.20$, respectively (from top to bottom). } \label{fig1}
\end{figure}
\end{center}

We show the behaviors of the DM growing mode solutions in Fig. \ref{fig1}. In Fig. \ref{fig1}a, we show the evolutions of growing mode solutions of $\delta_{\DM}$ for the different dark energy models when $\Omega_{\DM} = 0.23$. The dotted, solid and dashed lines correspond to $\omega_{de} = -1.5, -1.0$ and $-0.8$, respectively. If there is the same amount of the dark matter at the present epoch for the different dark models, then there will be more matter component in the past for the smaller value of $\omega_{de}$. Thus, dark energy model with the smaller value of $\omega_{de}$ maintains the longer linear growth behavior of $\delta_{\DM}$. We show the $\Omega_{\DM}^{0}$ dependence of $\delta_{\DM}$ for the $\Lambda$CDM model ({\it i.e.} with $\omega_{de} = -1.0$) in Fig. \ref{fig1}b. The dotted, solid and dashed lines correspond to $\Omega_{\DM}^{0} = 0.26, 0.23$ and $0.20$, respectively. Again if there is more dark matter at present epoch for the same models, then it maintains the longer linear growth behavior. The $\delta_{\DM}$ behaviors for the different dark models in Figs \ref{fig1}a look quite similar to $\delta_{\DM}$ of the cosmological constant model with the different values of $\Omega_{\DM}^{0}$ in \ref{fig1}b. It means that we have ambiguities in probing dark energy models with only using the growth of the DM.

In addition to this, we are able to find the exact analytic solution for the growth of the linear perturbation of baryon $\delta_{\B}$ from the above equation (\ref{dadeltaB}). The solution of $\delta_{\B}$ consists of two components because the evolution equation of $\delta_{\B}$ is an inhomogeneous second order differential equation.  Thus, we can separate the solution into the homogeneous solution and the inhomogeneous (particular) one
\ba && \delta_{\B} = \delta_{\B}^{\h} + \delta_{\B}^{\i} \, , \label{deltaBhi} \\
&& \fr{d^2 \delta_{\B}^{\h}}{da^2} + \Biggl( \fr{d \ln H}{d a} + \fr{3}{a} \Biggr) \fr{d \delta_{\B}^{\h}}{d a} = 0 \, , \label{dadeltaBh} \\ && \fr{d^2 \delta_{\B}^{\i}}{da^2} + \Biggl( \fr{d \ln H}{d a} + \fr{3}{a} \Biggr) \fr{d \delta_{\B}^{\i}}{d a} = \fr{4 \pi G \rho_{\DM}}{(aH)^2} \delta_{\DM} \, . \label{dadeltaBi}\ea The equation of the inhomogeneous solution $\delta_{\B}^{\i}$ is identical with that of $\delta_{\DM}$ in Eq. (\ref{dadeltaDM}). Thus, the inhomogeneous solution of $\delta_{\B}$ is nothing but $\delta_{\DM}$. We only need to solve the homogenous equation to obtain a complete solution of $\delta_{\B}$. $\delta_{\B}^{\h}$ is obtained from the similar method used in the previous works \cite{SK}. \be \delta_{\B}^{h} (Y) = c_{\h 1} \Bigl( \fr{Y}{Q} \Bigr)^{\fr{3 \omega_{\de} -1}{6 \omega_{\de}}} F \Bigl[ \fr{1}{2}, \fr{1}{2} - \fr{1}{6 \omega_{\de}}, \fr{3}{2} - \fr{1}{6 \omega_{\de}}, - Y \Bigr] + c_{\h 2} \, ,\label{deltaBh} \ee where $c_{\h 1}$ and $c_{\h 2}$ are the integral constants. However, we are not interested in the constant solution and remove $c_{\h 2}$ from the solutions. Thus the complete set of linear perturbation of the baryon is given by \be \delta_{\B} (a) \equiv b(a) \delta_{\DM} = c_{\h 1} a^{\fr{3 \omega_{\de} -1}{6 \omega_{\de}}} F \Bigl[ \fr{1}{2}, \fr{1}{2} - \fr{1}{6 \omega_{\de}}, \fr{3}{2} - \fr{1}{6 \omega_{\de}}, - Q a^{3 \omega_{\de}} \Bigr] + \delta_{\DM}(a) \, , \label{deltaBa} \ee where $b(a)$ is the baryon bias factor. Again, the analytic solution of $\delta_{\B}(a)$ in Eq. (\ref{deltaBa}) does not have any physical meaning before we fix the coefficient $c_{\h 1}$. Thus, we use the fact that $\delta_{\B} \simeq 0$ at decoupling epoch $a_{\dec} \sim 0.001$. Actually, we use $a_{\dec} = 0.01$ in order to include more relevant modes and more accurate to the matter dominated epoch approximation. Then we are able to obtain $c_{\h 1}$ from \be \delta_{\B}(a_{\dec}) = \delta_{\B}^{\h}(a_{\dec}) + \delta_{\DM}(a_{\dec}) = 0 \, . \label{inih} \ee We obtain the baryon bias factor in any epoch from the above equation (\ref{deltaBa}) \be b(a) = 1 + \fr{c_{\h 1} a^{\fr{3 \omega_{\de} -1}{6 \omega_{\de}}} F \Bigl[ \fr{1}{2}, \fr{1}{2} - \fr{1}{6 \omega_{\de}}, \fr{3}{2} - \fr{1}{6 \omega_{\de}}, - Q a^{3 \omega_{\de}} \Bigr]}{\delta_{\DM}(a)} \, . \label{ba} \ee The bias factor in Eq. (\ref{ba}) is usually expressed by the parameter as in the reference \cite{Padmanabhan} \be b(a) = 1 - \fr{c}{a} \, , \label{bap} \ee where $c$ is a constant to be fixed by $c = a_{\dec}$.
\begin{center}
\begin{figure}
\vspace{1.5cm}
\centerline{
\psfig{file=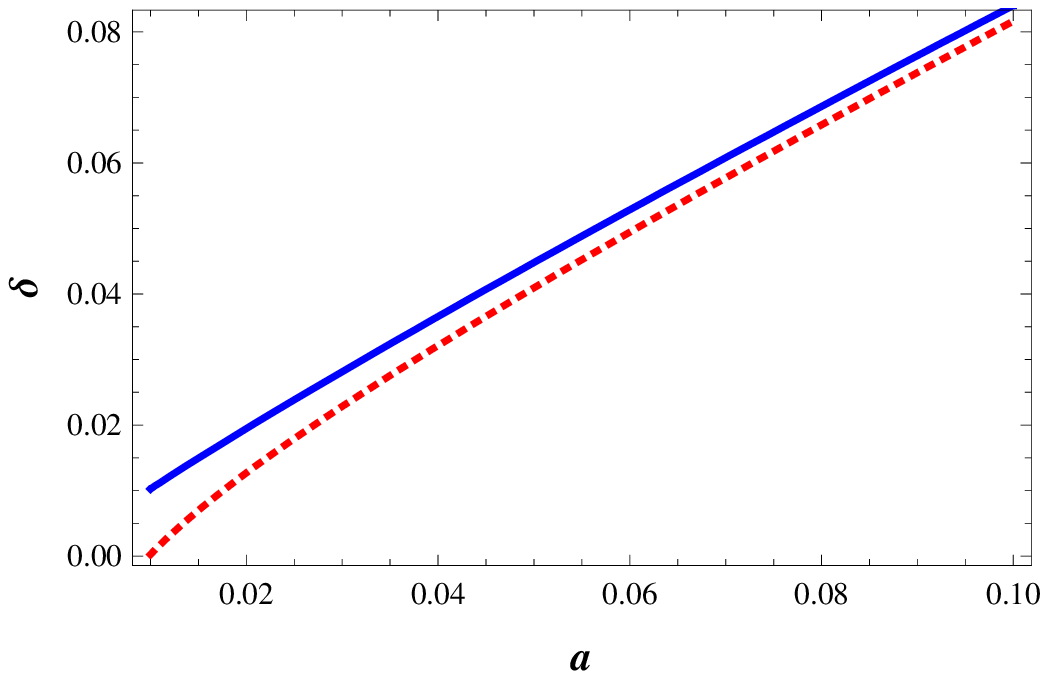, width=6cm} \psfig{file=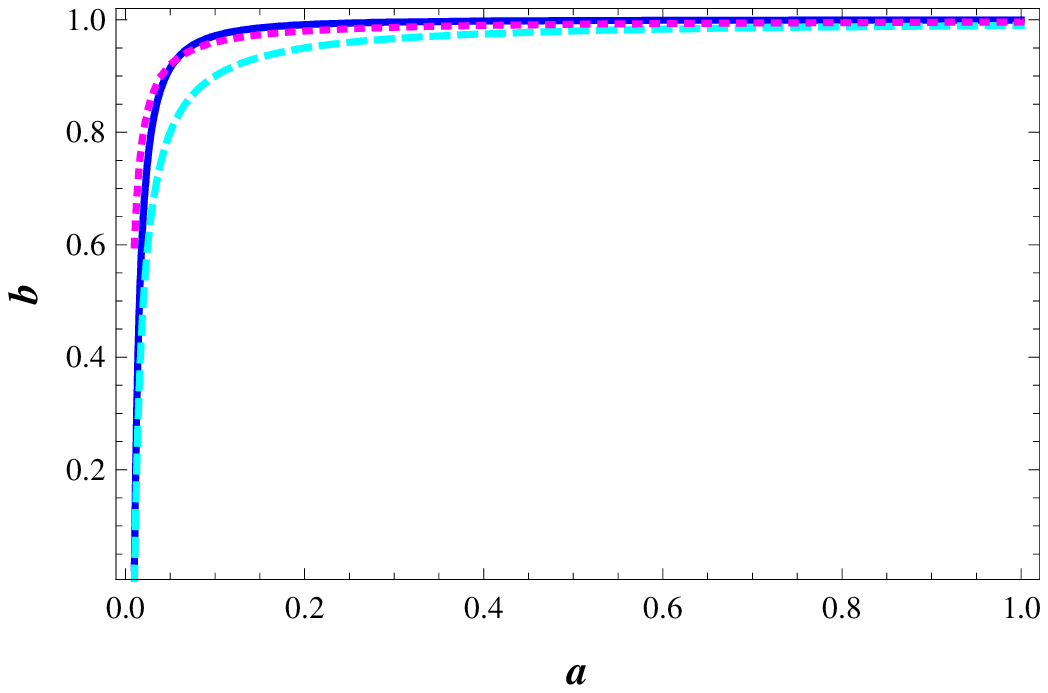, width=6cm} }
\vspace{-0.2cm}
\caption{ a) Evolutions of the perturbations of dark matter and baryon when $\Omega_{\DM}^{0} = 0.23$ and $\Omega_{b}^{0} = 0.04$ for the cosmological constant models $\omega_{de} = -1.0$ (from top to bottom).  b) Evolutions of bias factors $b(a)$ from the exact solutions and the parameterization of it with $c = 0.004$ and $0.01$ (from top to bottom). } \label{fig2}
\end{figure}
\end{center}
We show the evolutions of $\delta_{\DM}(a)$ (solid line) and $\delta_{\B}(a)$ (dotted line) when $\omega_{de} = -1.0$ in Fig \ref{fig2}a. We use $\Omega_{\DM}^{0} = 0.23$ and $\Omega_{\B} = 0.04$ in this figure. $\delta_{\B}(a)$ catches up with $\delta_{\DM}$ as expected. In Fig \ref{fig2}b, we also show the evolutions of baryon bias factors for the different cases. The solid lines correspond to the exact form of $b(a)$s for $\omega_{de} = -0.8, -1.0$ and $-1.5$ obtained from Eq. (\ref{ba}). All of the lines for the corresponding models are overlapped as one line. The dotted and dashed lines correspond to $c = 0.004,\, 0.01$, respectively when we use the approximate parameterization of $b(a)$ given in Eq. (\ref{bap}). The bias factor gets close to $1$ after $a \simeq 0.1$ because $\delta_{\B}$ gets very close to $\delta_{\DM}$ from this epoch. This is model independent feature of $b(a)$ and thus the given approximation of $b(a)$ in Eq. (\ref{bap}) is well matched with all of models we choose. Thus, baryon bias factor might not be a good tool for probing the dark energy model except some exotic cases \cite{Amendola}.

We summarize the coefficients of $\delta_{\DM}$ and $\delta_{\B}^{h}$ used in Fig. \ref{fig1}a and Fig. \ref{fig2} in table \ref{table1}.
\begin{center}
    \begin{table}
    \begin{tabular}{ | c | c | c | c |}
    \hline
    $\omega_{de}$ & $c_{1}$ & $c_{2}$ & $c_{h} \times 10^{4}$ \\ \hline
    -0.8 & -0.783555 & 0.692420 & -1.99786  \\ \hline
    -1.0 & -0.542664 & 0.697915 & -1.69270  \\ \hline
    -1.5 & -0.312706 & 0.701161 & -1.22785  \\ \hline
    \end{tabular}
    \caption{$\omega_{de}$ is the eos of the dark energy. $c_{1}$ and $c_{2}$ are the coefficients of $\delta_{\DM}$ given in Eq. (\ref{deltaDME}). $c_{h1}$ is the coefficient of $\delta_{\B}^{h}$. We use $\Omega_{\DM}^{0} = 0.23, \, \Omega_{\B}^{0} = 0.04$, $\Omega_{\de}^{0} = 0.73$ and $a_{i} = 0.01$ to get these values.}
    \label{table1}
    \end{table}
\end{center}
\begin{center}
\begin{figure}
\vspace{1.5cm}
\centerline{
\psfig{file=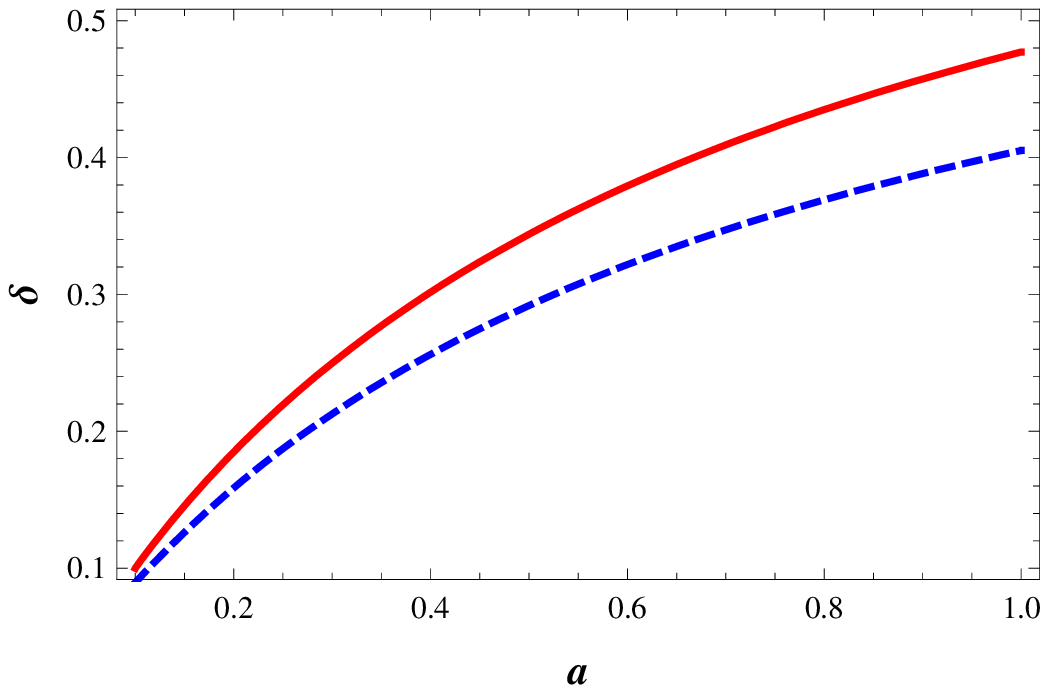, width=6cm} \psfig{file=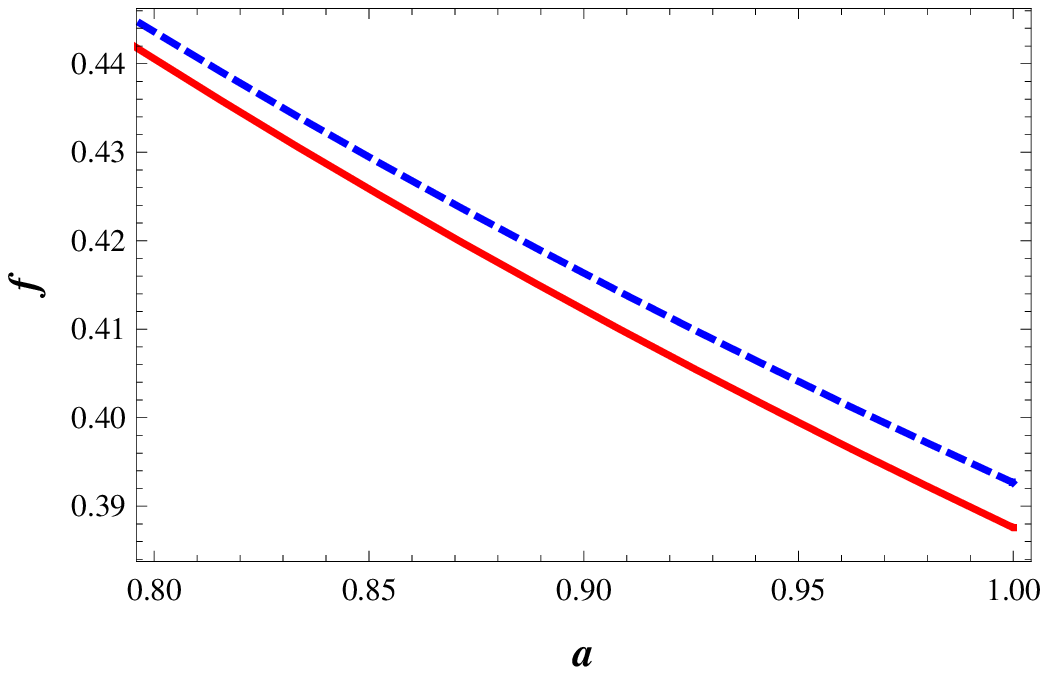, width=6cm} }
\vspace{-0.2cm}
\caption{  a) Evolutions of growth factors $D_{g}$ and $D_{\ws}$ (from top to bottom) when $\omega_{\de} = -0.4$ and $\Omega_{m}^{0} = 0.2$. b) Evolutions of $f(a) \equiv \Omega_{m}(a)^{\gamma}$ and $f$ (from top to bottom) for the same values of $\omega_{\de}$ and $\Omega_{m}^{0}$. } \label{fig3}
\end{figure}
\end{center}
We also emphasize that these solutions in Eq. (\ref{deltaDME}) and Eq. (\ref{deltaDM}) are different from the well known approximate solution in the reference \cite{Waga}. \be \delta_{g}^{\sw} = c_{\sw} \Biggl( \fr{\Omo}{\Odeo} \Biggr)^{\fr{1}{3\omde}} a F \Bigl[ -\fr{1}{3 \omde}, \fr{1}{2} - \fr{1}{2\omde}, 1 - \fr{5}{6 \omde}, -\fr{\Odeo}{\Omo} a^{-3\omde} \Bigr] \, . \label{Dsw} \ee The solution given in the reference \cite{Waga} is claimed to be a growing mode solution of Eq. (\ref{deltaDM}). However, one is not able to separate the correct growing mode solution with decaying mode one from the general solution except $\omega_{de} = -1/3$ or $-1$ \cite{Dodelson}. It is also easy to check from the initial conditions of the growing mode solution in Eq. (\ref{ini}). If one use the solution in Eq. (\ref{Dsw}), then one is not able to obtain the consistent coefficient value $c_{\sw}$ from the two initial conditions. We also need to interpret the solution in the correct way \cite{SK2}. The behavior of the exact analytic solution is consistent with the numerical result in the reference \cite{Dodelson}. Furthermore, the exact solutions are useful to extend to the other general models like modified theory of gravities and apply for the matter power spectrum, the mass function, the large scale statistics and so on. Also from the exact solution we are able to easily investigate the dependence of physical quantities on both model $\omega_{\de}$ and the present matter energy density contrast $\Omega_{m}^{0}$ \cite{SK}. This fact is easy to be ignored if one uses the well known fitting formulae \cite{WS}.

For the comparison of the exact solution with the fitting formulae, we use the total matter density growth factor instead of the DM growth factor given in Eq. (\ref{deltaDM}). The commonly used approximation of growth index parameter $\gamma_{ws}$ is given by \cite{WS} \be \gamma_{ws} \simeq \fr{3 (1 - \omega_{de})}{5- 6 \omega_{de}} + \fr{3}{125} \fr{(1-\omega_{de})(1-3\omega_{de}/2)}{(1-6\omega_{de}/5)^3} ( 1 - \Omega_{m}(a)) \, . \label{gammaWS} \ee The variation of $\gamma_{ws}$ with respect to $a$ is negligible. From this approximate parametrization $\gamma_{ws}$, the growth factor $\delta_{\ws}(a)$ is obtained \cite{WS,Linder} \be \delta_{ws}(a) \simeq a \exp \Biggl[ \int_{0}^{a} \fr{da'}{a'} \Bigl(\Omega_{m}(a')^{\gamma_{ws}} - 1 \Bigr) \Biggr] \, . \label{Da} \ee The above expression is slightly different from the original $\delta_{ws}$ in \cite{WS} which is normalized to the growth factor today. It is known that the value of the growth factor and the growth index obtained from the above fitting formulae are very close to the correct ones when $\omega_{de}$ is close to $-1$ \cite{Linder}. $\delta_{ws}$ and $f_{sw}$ are good approximations for small $\omega_{de}$ and big $\Omega_{m}^{0}$. However, we need to investigate the validity of this fitting formula for general cases. As a counterexample, we depict the evolutions of the growth factor $\delta$ ({\it i.e.} $D$) and the growth index $f(a) = \fr{d \ln \delta}{d \ln a} \equiv \Omega_{m}(a)^{\gamma}$ for $\omega_{de} = -0.4$ model with $\Omega_{m}^{0} = 0.2$ in Fig. \ref{fig3}. We show the evolution of the exact growth factor $\delta_{g}$ and that of the approximate one $\delta_{ws}$ in the left panel of Fig. \ref{fig3}. The solid line and the dashed line correspond to $\delta_{g}$ and $\delta_{ws}$ when $\omega_{de} = -0.4$ and $\Omega_{m}^{0} = 0.2$. The error of the present value of the growth factor is about $17$ \% in this model. We also show the evolution of the growth index $f_{ws}$ and $f$ for the same model in the right panel of Fig. \ref{fig3}. The dashed line and the solid one correspond to $f_{ws}$ and $f$, respectively. The errors of the growth index for the entire epoch are smaller than those of the growth factor. It is about $5$ \% only at the present epoch.

Thus, even though it is known as that the fitting formula is good for the cosmological concordance model, one should be careful for extending this formula to general models with other values of cosmological parameters. One may solve the physical quantities numerically without worrying about those case. However, it is useful and straight forward to use the exact solution if we are able to extend the exact solution to the general models like the time varying $\omega_{de}$ \cite{SK2}. We also are able to apply the exact analytic solution to probe the matter power spectrum, the evolution of mass function, the gravitational lensing and etc.

We thanks S.~Nesseris and I.~Waga for useful comment.

\end{document}